\documentclass[twocolumn,prl,showpacs,preprintnumbers,amssymb]{revtex4-1}

\usepackage{epsfig}
\usepackage{dcolumn}
\usepackage{bm}
\usepackage{hyperref}

\def\IC{\bf C}
\def\IZ{\bf Z}

\def\z2z2{$\IC^3/(\IZ_2\times\IZ_2)$}

\def\id{{\bf 1}}

\def\cG{\cal G}

\def\cL{{\cal L}}

\def\cO{\cal O}

\def\cR{\cal R}

\def\cp{\mbox{\bbbold C}\mbox{\bbbold P}}

\def\a{\alpha}

\def\b{\beta}

\def\d{\delta}\def\D{\Delta}

\def\k{\kappa}
\def\l{\lambda}

\def\p{\pi}

\def\s{\sigma}

\def\th{\theta}

\def\beq{\begin{equation}}\def\eeq{\end{equation}}
\def\beqa{\begin{eqnarray}}\def\eeqa{\end{eqnarray}}
\def\barr{\begin{array}}\def\earr{\end{array}}

\def\wt{\widetilde}

\def\ds {{\del \hspace{-6.4pt} \slash}\;}

 \let\br=\bigr

\def\bd{\begin{document}}
\def\ed{\end{document}}
\def\ba{\begin{array}}
\def\ea{\end{array}}
\def\bea{\begin{eqnarray}}
\def\eea{\end{eqnarray}}
\def\ft#1#2{{\textstyle{{\scriptstyle #1}\over {\scriptstyle #2}}}}
\def\fft#1#2{{#1 \over #2}}
\newcommand{\be}{\begin{equation}}
\newcommand{\ee}{\end{equation}}
\newcommand{\eq}[1]{(\ref{#1})}
\def\eqs#1#2{(\ref{#1}-\ref{#2})}
\def\det{{\rm det\,}}
\def\tr{{\rm tr}}
\newcommand{\ho}[1]{$\, ^{#1}$}
\newcommand{\hoch}[1]{$\, ^{#1}$}
\def\ra{\rightarrow}

\def\Xh{\hat{X}}
\def\ah{\hat{a}}
\def\xh{\hat{x}}
\def\yh{\hat{y}}
\def\ph{\hat{p}}
\def\G{{\cal G}}
\def\Dth{{\Delta_\th}}

\def\bk{{\bf k}}
\def\bx{{\bf x}}
\def\br{{\bf r}}
\def\tr{{\rm tr \,}}
\def\Tr{{\rm Tr \,}}
\def\diag{{\rm diag \,}}
\def\tg{{\rm tg \,}}
\def\NPB#1#2#3{Nucl. Phys. B {\bf #1} (19#2) #3}
\def\PLB#1#2#3{Phys. Lett. B {\bf #1} (19#2) #3}
\def\PLBold#1#2#3{Phys. Lett. {#1B} (19#2) #3}
\def\PRD#1#2#3{Phys. Rev. D {\bf #1} (19#2) #3}
\def\PRL#1#2#3{Phys. Rev. Lett. {\bf #1} (19#2) #3}
\def\PRT#1#2#3{Phys. Rep. {\bf #1} C (19#2) #3}
\def\MODA#1#2#3{Mod. Phys. Lett.  {\bf #1} (19#2) #3}
\def\ov{\overline}

\def\preal{{\rm Re\,}}
\def\pim{{\rm Im\,}}

\def\ds{\displaystyle}
\def\yzero{\smash{\hbox{$y\kern-4pt\raise1pt\hbox{${}^\circ$}$}}}
\def\p{\partial}
\def\a{\alpha}
\def\b{\beta}
\def\g{\gamma}
\def\d{\delta}
\def\beq{\begin{equation}}
\def\eeq{\end{equation}}
\def\beqa{\begin{eqnarray}}
\def\eeqa{\end{eqnarray}}
\def\Om{\Omega}
\def\om{\omega}
\def\th{\theta}
\def\vt{\vartheta}
\def\vphi{\varphi}
\def\-{\hphantom{-}}
\def\ov{\overline}
\def\s2{\frac{1}{\sqrt2}}
\def\wh{\widehat}
\def\wt{\widetilde}
\def\oh{\frac{1}{2}}
\def\tr{{\rm tr \,}}
\def\Tr{{\rm Tr \,}}
\def\diag{{\rm diag \,}}
\def\vac{|0 \rangle}
\def\vm{\relax{n_V}}
\def\cc{{\cal C}}
\def\ck{{\cal K}}
\def\ci{{\cal I}}
\def\cu{{\cal U}}
\def\cg{{\cal G}}
\def\cn{{\cal N}}
\def\cam{{\cal M}}
\def\cp{{\cal P}}
\def\ct{{\cal T}}
\def\cv{{\cal V}}
\def\cz{{\cal Z}}
\def\ch{{\cal H}}
\def\cf{{\cal F}}
\def\tv{\tilde v}
\def\Dsl{\,\raise.15ex\hbox{/}\mkern-13.5mu D} 
\def\IZ{Z\kern-.4em  Z}
\def\id{{\rm 1}}

\def\ti{\times}
\def\til{\tilde}
\def\eps{\epsilon}
\def\k{\kappa}
\def\A{\Arrowvert}
\def\cw{{\cal W}}
\def\G{\Gamma}
\def\car{{\cal R}}
\def\l{\lambda}
\def\raw{\rightarrow}
\def\Raw{\Rightarrow}
\def\inte{{\bf Z}}
\def\cpx{{\bf C}}
\def\real{{\bf R}}
\def\Lam{\Lambda}
\def\D{\Delta}
\def\cb{{\cal B}}
\def\ca{{\cal A}}

\begin{document}

\preprint{MAD-TH-13-01}

\title{Milli-Charged Dark Matter in Quantum Gravity and String Theory}

\author{Gary Shiu$^{1,2}$, Pablo Soler$^{1,2}$, and Fang Ye$^{1,2}$}
\affiliation{\small\slshape  $^{1}$ Department of Physics, 1150 University Avenue, University of Wisconsin, Madison, WI 53706 \\
$^{2}$ Department of Physics and Institute for Advanced Study, Hong Kong University of Science and Technology, Hong Kong}
\begin{abstract}
We examine the milli-charged dark matter scenario from a string theory perspective. In this scenario, kinetic and mass mixings of the photon with extra $U(1)$ bosons are claimed to give rise to small electric charges, carried by dark matter particles, whose values are determined by continuous parameters of the theory. This seems to contradict {\it folk theorems} of quantum gravity that forbid the existence of irrational charges in theories with a single massless gauge field. By considering the underlying structure of the $U(1)$ mass matrix that appears in type II string compactifications, we show 
 that milli-charges arise exclusively through kinetic mixing, and require the existence of at least two exactly {\it massless} gauge bosons.
\end{abstract}
\pacs{95.35.+d, 11.25.w}
\maketitle

The quest for understanding the nature of dark matter (DM) continues to be an inspiration for 
new  physics scenarios.
Cosmological observations provide compelling evidences that 
a substantial fraction of our universe is made up of DM that cannot be composed of any of the known particles.
At the same time, attempts to understand electroweak symmetry breaking (EWSB) also invariably require new particles beyond the Standard Model (BSM).
The hierarchy problem which 
lies at the heart of EWSB
significantly highlights the sensitivity to high scale physics.
Thus it is worthwhile to
examine whether the new particles introduced
in BSM (or perhaps related DM) scenarios
are motivated from the perspective of a fundamental theory such as string theory.

Besides supersymmetry, axions and an extended gauge sector are among the most pervasive  
elements in string constructions.
While extra $U(1)$ gauge symmetries are
common
 in bottom-up scenarios, those
 appearing in
string models exhibit further interesting features.
The aim of this work is to examine the implications of extra $U(1)$ symmetries in string theory
for the milli-charged DM scenario \cite{FCHAMPS}. This scenario postulates the existence of BSM particles with tiny electric charges.
While the observed particles appear to carry quantized electric charges, the DM particles do not need to do so, as long as the electric charge is small enough to evade observational bounds \cite{Davidson:2000hf}.  
One way to obtain effective milli-charged particles, even if the hypercharge $Y$ is {\it a priori} quantized, 
is to introduce an extra $U(1)'$ whose kinetic mixing with $U(1)_Y$ is such that a particle charged under $U(1)'$ appears to have a small (generically irrational) coupling to the photon \cite{Holdom:1985ag}.
It was further suggested that this new $U(1)'$ 
could be a massive $Z'$ \cite{Feldman:2007wj} and thus if the masses and couplings were in the right range, the $Z'$ boson and the milli-charged particles could be detected through a confluence of astrophysical and collider experiments \cite{Cheung:2007ut}.

On the other hand, general {\it folk theorems} involving black holes 
suggest
that theories with a single massless gauge boson cannot be consistently coupled to quantum gravity if there exist matter fields with irrational charges \cite{Banks:2010zn}. 
Given the somewhat conjectural and abstract nature of these statements, we
shall investigate
 whether the milli-charged 
DM
 scenario can indeed be realized in a theoretically  motivated framework. 

To concretely illustrate our point,
we focus on type IIA string compactifications to four dimensions with intersecting D6-branes \cite{Blumenhagen:2005mu}.
As we will see, the underlying structure of the mass matrix of gauge bosons in those theories has important consequences regarding charge quantization. If the theory contains only one exactly massless $U(1)$ gauge boson (the photon), electric charges are quantized, and models with very small charges are barely realizable. On the other hand, if there exists an extra exactly massless $U(1)$ (a {\it dark photon}), milli-charges may arise by its kinetic mixing with the photon as in \cite{Holdom:1985ag}.

\vspace{.4cm}

Consider
 the  
Lagrangian for $N$ $U(1)$'s, arranged in a vector $\vec{A}^{\,T}=(A_1,\ldots,A_N)$, with 
 general
 kinetic mixing matrix $f$, mass matrix $M$, and interactions with matter:
\beq\label{lag}
\cL=-\frac{1}{4}\vec{F}^{\,T}\cdot f\cdot\vec{F}-\frac{1}{2} \vec{A}^{\,T}\cdot M^2\cdot\vec{A}+\overline{\psi}\left(i\partial\!\!\!/ +  \vec{q}_{\psi}^{\,T}\cdot\vec{A}\!\!\!/ \right)\psi
\eeq
The charges $\vec{q}_\psi$ are assumed to be quantized (as happens in string theory), and we use a normalization such that they are all integral. The mass matrix $M^2$ may arise either from the Stueckelberg or the Higgs mechanism. 

To facilitate our analysis, we may go to a basis in the space of $U(1)$'s in which both the kinetic and the mass terms are diagonal. This can be done in three steps: 
\begin{list}{\labelitemi}{\leftmargin=1em}
\item 
 First, perform an orthogonal transformation $\vec{A}\to {\cO}\cdot{\vec A}$ to bring the kinetic term to a diagonal form 
\beq\label{kineticdiag}
{\cO}^{\,T}\cdot f \cdot{\cO}={\mathrm{diag}}(\,g_1^{-2}~\ldots~g_N^{-2})\equiv \Lambda^{-2}
\eeq

\item Next, reabsorb the coupling constants by applying the matrix $\Lambda$ to the gauge fields, so that the resulting kinetic matrix is just the identity.

\item Finally, diagonalize the resulting mass matrix with a second orthogonal transformation $\cR$ such that 
\beq\label{R-matrix}
 {\cR}^{\,T}\cdot\Lambda\cdot{\cO}^{\,T}\cdot M^2\cdot{\cO}\cdot{\Lambda}\cdot{\cR}={\mathrm{diag}}(\,m_1^{2}~\ldots~m_N^{2})
\eeq
The values $m_i$ are the masses of the physical gauge bosons (those that propagate without mixing).
Some
of them may be zero, as will happen if the rank of the matrix $M^2$ is lower than $N$.
\end{list}

After the diagonalization process, the vector of gauge bosons $\vec{A}'$ in the final basis will be given by
\beq\label{vectors}
\vec A={\cO}\cdot{\Lambda}\cdot{\cR}\cdot{\vec A}'
\eeq
By inserting this expression in the original Lagrangian (\ref{lag}) we see that the couplings of the matter fields to these bosons will be parameterized by 
\beq\label{charges}
{\vec{q}\,'}^{T}=\vec{q}^{\,T}\cdot{\cO}\cdot{\Lambda}\cdot{\cR}
\eeq
In general, the transformation matrices $\cO$, $\Lambda$ and $\cR$ depend on several continuous parameters of the theory, such as coupling constants and masses, and they do so in a very complicated way. It is expected that, through (\ref{charges}), such dependence is transmitted to the 
matter charges, which in general will not be quantized.

This mechanism has been used repeatedly in the literature to generate scenarios in which 
DM
 particles carry a small (generically irrational) electric charge.
This lack of charge quantization, however,
seems to clash with black hole arguments that a theory with irrational charges cannot be consistently coupled to quantum gravity \cite{Banks:2010zn}.

In the rest of this letter we will see how such problems are solved in models arising from string theory. To be concrete, we shall 
make our arguments in the context of intersecting D6-brane models in type IIA string compactifications, though
we expect that our results can be applied (e.g. through dualities) to many other stringy constructions, and even 
more
 generally to any theory consistently coupled to quantum gravity.

\vspace{0.15cm}
In type IIA model building, the Standard Model (SM) gauge group 
is realized by open strings living on the worldvolumes of D6-branes that span the four non-compact Minkowski dimensions, and wrap three-cycles of an internal six-dimensional compactification space $X$. In general, for a stack of $n$ such branes, the gauge group is locally $U(n)\cong U(1)\times SU(n)$, which contains an abelian factor that is the main object of our analysis \cite{group}.

In general, several such stacks are needed to reproduce the gauge group of the SM, and even more appear in hidden sectors which are often needed to satisfy tadpole cancellation conditions of the compactification. Hence, several $U(1)$ factors, either coupled or not to the SM particles, arise generically in such models. Their low energy Lagrangian is described by (\ref{lag}), with 
mass matrix coming either from the Stueckelberg or the Higgs mechanism.

\nopagebreak

In the Stueckelberg case, the mass matrix comes from the coupling of the gauge bosons to axionic scalar fields $\phi^i$ that arise from reducing the Ramond-Ramond (RR) three-form on three-cycles of the internal space:
\beq\label{stueckelberg}
\cL_{\mathrm{St}}\sim \sum_{i,j} {\cG}_{ij}(\partial_\mu\phi^i+k^i_a A^a_\mu)(\partial^\mu\phi^j+k^j_b A^{b\,\mu})
\eeq
where the indices $i,j$ run over all the RR axions (i.e. over the homology of three-cycles of the internal space $X$). The matrix ${\cG}$ is the (positive definite) metric of the complex structure moduli space. This metric depends in a complicated way on continuous parameters of the theory (the moduli). Fortunately, its exact form will not be important for our arguments.
What
will play a crucial role are the {\it integer} intersection numbers
 $k^i_a=[\Pi_a]\cdot[\Pi^i]$
  between the three cycle $[\Pi_a]$ wrapped by the $a$-th stack, and the three-cycle $[\Pi^i]$ associated to the RR axion $\phi^i$.

The Higgs mechanism can be written in exactly the same way, by expressing the Higgs fields $\rho^j=|\rho^j|e^{i\alpha^j}$ in terms of their absolute values and phases. 
 Here,
 the matrix $\cG$ would encode their vacuum expectation values ${\cG}_{ij}\sim\langle|\rho^i|\rangle\,\delta_{ij}$. The role of the axions $\phi^i$ would be played by the phases $\alpha^i$, and the numbers $k^i_a$ 
 would 
 correspond to the charges of $\rho^i$ under $U(1)_a$. 
 These are also integers, in fact $k^i_a\in\{0,\pm1,\pm2\}$, depending on how the ends of the corresponding open string attach to the brane $[\Pi_a]$.

  If 
the Higgs is also charged under non-abelian gauge groups, an extra gauge field must be included in the vector $\vec A$, corresponding to the component of the non-abelian group that mixes with the abelian factors (e.g. the third component of the isospin $SU(2)_L$ in the usual EWSB of the SM). Hence, our arguments will apply to hypercharge as well as to electromagnetism itself after EWSB.

Summarizing,
 the mass matrix 
 for gauge fields 
 arising
 in string compactifications, including both the Stueckelberg and the Higgs mechanisms, can be written 
 as
\beq\label{structure}
(M^2)_{ab}={\cG}_{ij} k^i_a k^j_b\,, ~~~~~~(M^2=K^{\,T}\cdot {\cG}\cdot K)\,.
\eeq
The positive-definite matrix $\cG$ depends on continuous parameters of the theory, while the matrix $K$ has {\it integer} entries that encode intersection numbers of the compactification and charges of the Higgs fields. The integrality of this matrix is the key point of the following discussion.

\vspace{.2cm}
Now, how does the
underlying structure (\ref{structure}) of the mass matrix 
reflect in 
 the transformation matrices $\cO$, $\Lambda$ and $\cR$ of eq.~(\ref{charges})?
In general, these 
matrices
 still depend in a complicated manner on the continuous parameters of the theory because the kinetic matrix $f$ and the matrix $\cG$ do. However, 
as we now discuss,
the case of massless gauge bosons 
and their couplings
 is special.

Notice
 that the columns of the matrix $\cR$ involved in the diagonalization of the mass matrix (see eq.~(\ref{R-matrix})) are just a set of orthonormal eigenvectors of the matrix $(\Lambda\cdot{\cO}^{\,T}\cdot M^2 \cdot{\cO}\cdot\Lambda)$. 
Let us construct 
some of these vectors.

\vspace{.4cm}
\noindent {\bf A single massless boson:} First
consider the case 
where
the matrix $K$ has rank $N-1$, so 
 there exists a unique $N$-vector $\vec{v}$ such that $K\cdot\vec{v}=0$. This vector encodes the linear combination of gauge bosons that remains massless after
 diagonalization.
 Equivalently, in type IIA language, this is the linear combination of 
 three-cycles 
 wrapped by branes that is trivial in
 the
  homology of the internal space X. 
  Such 
  vector ${\vec v}$ can always be found with integer entries.

Now, $\vec v$ is an eigenvector of the original mass matrix $M^2$ with zero eigenvalue. It is then straightforward to construct an eigenvector $\vec{v}\,'$ (also with zero eigenvalue) of the transformed matrix  $\Lambda\cdot{\cO}^{\,T}\cdot M^2 \cdot{\cO}\cdot\Lambda$. It is simply 
\beq\label{vector}
\vec{v}\,'\equiv \Lambda^{-1}\cdot {\cO}^{\,T}\cdot \vec v
\eeq
Hence, the orthogonal matrix $\cR$ will take the form
\beq
\cR=
\left(
\begin{array}{ccc}
&&\\
\frac{\vec{v}\,'}{|\vec{v}\,'|}&\tilde{\cR}& \\
&&\\
\end{array}\right)\,,
\eeq
where $\tilde{\cR}$ is a complicated moduli dependent reduced matrix that encodes the massive gauge bosons. 
 Substituting
 these expressions into (\ref{charges}), we can read off
 the couplings of matter to the physical massless boson associated to $\vec v$:
\beq
q'_{\vec{v}} =\frac{1}{|\vec{v}\,'|} \vec{q}^{\,\,T}\cdot{\cO}\cdot\Lambda\cdot\vec{v}\,'= \frac{1}{|\vec{v}\,'|}\, \vec{q}^{\,\,T} \cdot \vec{v}
\eeq
Notice 
that all continuous parameters in this expression are encoded in 
the 
overall prefactor $1/|{\vec{v}'}|$, which can be identified with the gauge coupling constant.
The charges
are still integral  since 
 both 
  $\vec q$ and $\vec v$ have integer entries.

Hence, we see that in models in which all $U(1)$ bosons but the photon gain a mass (either Stueckelberg or Higgs), the electric charges of the theory are quantized. This means that in such setups, DM can only carry fractional charges with respect to that of the electron, and these are not tunable by continuous parameters. 

Furthermore, 
the smallness of these fractional charges as 
required to avoid conflict with experiment (usually taken to be at least $<10^{-2}$ for sufficiently low FCHAMP masses), are not realized in any known realistic string compactification. One would need large integral entries of the vector $\vec{v}$ 
to get such suppressions. However, 
$\vec{v}$ just expresses the linear combination of cycles wrapped by branes that is trivial in the homology of the internal space, and its entries are always of order ${\cO}(1)$. As an example, the popular SM-like {\it Madrid models} of \cite{Ibanez:2001nd} contain three $U(1)$ bosons, and the linear combination of them that remains massless (the hypercharge) is encoded in the vector $\vec{v}=(1,-3,3)$. Higher numbers (i.e. smaller fractional charges) require large wrapping numbers (at least ${\cO}(100)$), which are difficult to implement in consistent (tadpole-free) and realistic string compactifications.

\vspace{.4cm}

\noindent{\bf Two massless bosons and milli-charges:}
We have 
 seen that milli-charges do not arise in models in which only the SM photon remains massless. Hence, to 
realize
 the milli-charged scenario, we need to
 consider
 setups
 where at least an extra gauge boson (a {\it hidden photon}) remains exactly massless. 
The reasoning is similar to that
for the single massless boson case.
Consider 
the
case in which the 
$K$ matrix of eq.~(\ref{structure}) 
has 
rank $N-2$. Generalizations  
with more massless $U(1)$'s 
are
 straightforward.

In this case, we can easily find two linearly independent vectors $\vec{v}_1$ and $\vec{v}_2$, again with integer entries, such that $K\cdot \vec{v}_{1,2}=0$. At this point we have the freedom to choose any two of them (they do not even need to be orthogonal). 

Once more, we can construct out of these vectors new eigenvectors ${\vec{v}\,'}_{1,2}$ of the matrix
$\Lambda\cdot{\cO}^{\,T}\cdot M^2 \cdot{\cO}\cdot\Lambda$
\begin{equation}\label{2v'}
{\vec{v}\,'}_{1,2}\equiv \Lambda^{-1}\cdot{\cO}^T\cdot\vec{v}_{1,2}.
\end{equation}
However, these will not in general be orthogonal to each other, so they will not correspond to columns of the orthogonal matrix $\cR$
that
 we are looking for. 

What we can do is project one of them, say $\vec{v}\,'_2$ onto the subspace orthogonal to $\vec{v}\,'_1$, i.e. we can define
\begin{eqnarray}\label{2v''}
\vec{v}\,''_2&\equiv&\vec{v}\,'_2-\frac{(\vec{v}\,'^{\,T}_2\cdot \vec{v}\,'_1)}{|\vec{v}\,'_1| ^{2}}\,\vec{v}\,'_1\\
&=&\Lambda^{-1}\,{\cO}^T\left( \vec{v}_2-\frac{\vec{v}^{\,T}_2\cdot f\cdot\vec{v}_1}{|\vec{v}\,'_1|^{2}}\vec{v}_1\right)\nonumber\,,
\end{eqnarray}
which still corresponds to a massless eigenstate.
In the second line we have used (\ref{kineticdiag}) to reintroduce the original kinetic matrix $f$. The vectors  ${\vec{v}\,'_{1}}$ and ${\vec{v}\,''_{2}}$ are now orthogonal, so they will be part of the orthogonal matrix 
\beq\label{2orth}
{\cR}=\left(
\begin{array}{ccc}
&&\\&&\\
\frac{\vec{v}\,'_1}{|\vec{v}\,'_1|}&\frac{\vec{v}\,''_2}{|\vec{v}\,''_2|}&\tilde{\cR} \\
&&\\&&
\end{array}\right)\,.
\eeq
Now, by substituting (\ref{2v''}) and (\ref{2orth}) back into (\ref{charges}), we obtain the general expression for the couplings of matter fields to the massless gauge bosons associated to $\vec{v}_{1,2}$:
\begin{eqnarray}\label{finalcharges}
q_1&=&\frac{1}{|\vec{v}\,'_1|}\,\vec{q}^{\,T}\cdot \vec{v}_1\\
q_2&=&\frac{1}{|\vec{v}\,''_2|}\,\vec{q}^{\,T}\cdot \left( \vec{v}_2-\frac{\vec{v}^{\,T}_2\cdot  f\cdot\vec{v}_1}{|\vec{v}\,'_1|^{2}}\vec{v}_1\right)\equiv\frac{1}{|\vec{v}\,''_2|}\,\vec{q}^{\,T}\cdot (\vec{v}_2-\epsilon \,\vec{v}_1)\nonumber
\end{eqnarray}
where we have defined the milli-charge shift 
\beq\label{epsilon}
\epsilon=\frac{\vec{v}^{\,T}_2\cdot f\cdot\vec{v}_1}{|\vec{v}\,'_1|^{2}}\,.
\eeq
Again, the moduli dependent factors $1/|\vec{v}\,'_{1}|$ and $1/|\vec{v}\,''_{2}|$ can be thought of as the gauge coupling constants. We see that $q_1$ couplings are quantized, while $q_2$ couplings are not. Irrational milli-charges proportional to $\epsilon$ involve only the kinetic mixing matrix $f$, and are independent of the mass matrix $\cG$. This scenario is reminiscent of the kinetic mixing setup originally considered in \cite{Holdom:1985ag}. 
Here, we show that 
its reverse is true: existence of milli-charges {\it necessarily} implies two or more {\it massless} gauge bosons.

Our result is consistent with the continuous process of turning on and off the Higgs-mechanism, since the couplings of matter to {\it massive} $U(1)$'s 
need not be quantized.

Of course, any choice of vector $\vec{v}_1$ is valid. This serves to avoid the clash of irrational charges with quantum gravity. In \cite{Banks:2010zn}, it was argued that the evaporation of black holes carrying a small irrational charge under a non-quantized massless $U(1)$ would lead to inconsistencies. In our setup, however, for any such object with a charge vector $\vec{q}_{BH}$, we can define a convenient basis of $U(1)$'s in which it couples to a single massless gauge boson with an integral charge, thus avoiding 
possible tension with black hole evaporation arguments.

\vspace{0.15cm}

Finally, let us present a viable 
setup in which a hidden sector 
dark photon $\tilde{\gamma}$ mixes kinetically with the SM photon $\gamma$ and gives rise to milli-charged particles. A useful choice of vectors ${\vec{v}_{1,2}}$ is one in which the SM particles are not charged at all under the hidden photon,
while the DM particles only carry milli-electric charges.  
If we define the vectors $\vec{v}_{1}$ and $\vec{v}_{2}$ of (\ref{finalcharges}) as those corresponding to the hidden and the normal photon, respectively, we can satisfy these conditions by imposing
\begin{eqnarray}\label{constraints}
q_{\tilde{\gamma}}^{SM}=0 ~~~~&\Longleftrightarrow&~~~~~ \vec{q}_{SM}^{\,\,T}\cdot\vec{v}_1=0\nonumber\\
q_\gamma^{DM}\propto \epsilon ~~~~&\Longleftrightarrow&~~~~~ \vec{q}_{DM}^{\,\,T}\cdot\vec{v}_2=0
\end{eqnarray}

The models we consider consist of two sectors. A visible one (V) that corresponds to the SM, and a 
Hidden one (H) which hosts the dark photon and the DM. We assume that both sectors are away from each other and do not intersect. The vector of abelian gauge bosons in this setup factorizes as
$\vec{A}=(\vec{A}_V,\vec{A}_H)$.
Accordingly, the kinetic mixing matrix that gives rise to the milli-charges takes the form
\beq\label{kinmix}
f=\left( \begin{array}{cc}
f_V & \chi \\
\chi^{\,T} & f_H\\
\end{array} \right)\,.
\eeq
Here, 
$\chi$ represents the kinetic mixing between the hidden and the visible sector gauge bosons.

We next require that, independently in each sector, a linear combination of the three-cycles wrapped by the branes is trivial in homology. That is, each sector hosts a massless $U(1)$. This means that the two vectors $\vec{v}_{1,2}$  annihilated by the integral matrix $K$ can be expressed as $\vec{v}_2 ^{\,T}=(\vec{v}_\gamma~\vec{0})$ and $\vec{v}_1 ^{\,T}=(\vec{0}~\,\vec{v}_{\tilde{\gamma}})$. Relations (\ref{constraints}) are then trivially satisfied because the two sectors do not intersect each other. However, 
the kinetic mixing terms $\chi$ of (\ref{kinmix}) generate milli-electric charges through (\ref{epsilon}) of order
\beq
\epsilon=\frac{(\vec{v}^{\,T}_\gamma~\ \vec{0}\,)\cdot\left( \begin{array}{cc}
f_V & \chi \\
\chi^{\,T} & f_H\\
\end{array} \right)\cdot
\left(\begin{array}{c}
\vec{0}\\ \vec{v}_{\tilde{\gamma}}
\end{array}\right)}{|\vec{v}\,'_1|^{2}}
=\frac{\vec{v}^{\,T}_\gamma\cdot\chi\cdot\vec{v}_{\tilde{\gamma}}}{|\vec{v}\,'_1|^{2}}
\eeq
Hence, we see that the milli-charge depends only on the off-diagonal blocks $\chi$ of the kinetic matrix. Some estimations for $\chi$ were obtained in \cite{Goodsell:2009xc} and some more explicit results are available for toroidal compactifications
\cite{Abel:2008ai}. In Calabi-Yau compactifications we may expect lower values of $\chi$ since one-cycles are generically absent, and $\chi$ is generated (in the closed string channel) by the exchange of string modes, rather than  
Kaluza-Klein modes.

\vspace{.14cm}

In summary, we have shown that milli-charged DM necessarily requires two or more {\it massless} $U(1)$ bosons. 
Previous works (e.g., \cite{Feldman:2007wj,Cheung:2007ut}) which invoked 
mixing
of the photon with massive $U(1)$'s to generate milli-charges, though phenomenologically rich, are either incompatible with quantum gravity, or else reduce to highly implausible FCHAMP models.
Undoubtedly, a strong motivation for considering Stueckelberg $Z'$ in these works is the potential DM-LHC connection. Massive $Z'$ bosons, if sufficiently light and coupled non-gravitationally to the SM, can lead to distinctive collider signatures \cite{Langacker:2008yv}.
Our results indicate that even if such massive $Z'$'s are found at the LHC, they are not the extra gauge fields responsible for milli-charged DM.
Nonetheless, Stucekelberg inspired DM scenarios without milli-charges \cite{Feldman:2011ms} remain a viable option to connect LHC physics and DM searches.

\vspace{.14cm}
\acknowledgments

We thank W.-Z.~Feng, M.~Goodsell, J.~Jaeckel, P.~Nath, N.~Piazzalunga, A.~Ringwald, A.~Uranga and C.~Weniger for discussions.
This work 
was supported in part by DOE grant DE-FG-02-95ER40896. GS and PS  
thank the 
University of Amsterdam for their 
hospitality.


\begin{thebibliography}{99}
\bibitem{FCHAMPS}
We distinguish milli-charges from fractionally charged (massive) particles (FCHAMPs). Charges are continuous parameters in the former, and discrete in the latter.
\bibitem{group}
{Non abelian factors (e.g. $SU(n)$, $SO(n)$) that also arise enter our arguments only through the Higgs mechanism.}
\end{thebibliography}
\end{document}